\RequirePackage{fix-cm}
\documentclass[onecolumn]{svjour3questa}       
\usepackage{amsmath,amssymb}
\usepackage{mathptmx}      
\usepackage[colorlinks=false]{hyperref}

\begin{document}

\title{Queues on interacting networks}

\author{Maria Vlasiou}

\institute{University of Twente , the Netherlands \&
Eindhoven University of Technology, the Netherlands.\\
          \email{m.vlasiou@tue.nl}
          }
\date{\today}

\maketitle

\section{Introduction}
Queues acting on networks is a time-tested topic \cite{harrison-BMSFS,kelly-RSN}. Queuing networks are very suitable to detect performance bottlenecks due to shared resources on distributed systems. To deal with common synchronisation mechanisms of distributed systems, Petri nets and their extensions have been analysed \cite{baccelli96,peterson-PNT}. Layered queuing networks have been proposed to deal with nested resources and structures \cite{DorsmanVlasiouBoxma,woodside02}. However, current studies in queuing theory do not incorporate interactions between networks in the performance analysis of the queuing system.

Many real-world systems have been modelled as graphs: a collection of nodes and their pairwise interactions. Despite being widespread, traditional networks often do not provide a faithful representation of reality. Rarely are single networks isolated; it is usually the case that several networks are interacting and interdependent on each other. In infrastructures, the banking systems are interdependent with the Internet and the electric power grid, and public transport is dependent on the power grid, which relies in turn on the water supply system to cool the power plants, etc. These are examples of a network of networks, i.e., networks formed by several interdependent networks. These `higher-order interactions' are better described by simplicial complexes and hypergraphs, i.e.\ more complex mathematical structures with respect to traditional graphs.

Nowadays, such types of networks are captured under the term \textit{interacting} networks, with e.g., multiplex networks being a special case \cite{bianconi-MN}. Their investigation has become ubiquitous in the last decade. This has revealed new patterns of interactions and functionality which arise from inherently high-order features and could not be understood by limiting the analysis of structural properties to pairwise links. From social contagion to synchronisation, the introduction of higher-order interactions in networked systems has already been shown to give rise to new emergent physical phenomena, which cannot be predicted by breaking higher-order interactions into low-order dyads.

\section{Problem statement}
My view is that queuing theory needs to incorporate high-order network interactions in the performance analysis of a queuing system. While random processes in static random structures are relatively well-understood, their analysis in dynamic interacting settings is still in its infancy. The dynamics of interacting networks cannot be captured with a single network with additional constraints. Interacting networks display phenomena that \textit{cannot be explained} by their individual elements: cascading failures in interdependent networks feature first-order transitions \cite{buldyrev10} and structurally sharp transitions \cite{darabi15}. Such transitions happen when building interconnected networks by blocks of independent networks \cite{radicchi13}. Interacting networks encode more information than single layers taken in isolation. Assumptions on the nature of the interconnections between network layers are overly simplistic \cite{critis}. Notably, it is unclear if `realistic' assumptions can be made concerning the interdependencies between the different layers \cite{bullmore09}. Their precise form is not well-understood, and if known at all, it may be impossible to describe.

I present some sample general questions below. A network of networks is formed by $M$ networks (layers), each formed by a number of nodes. All quantities may be infinite; in fact, the more realistic case of finite $M$ is a challenging problem. Examples of such networks are \cite{son12}, where each node on a layer depends on exactly one node on another layer, and \cite{bianconi14a}, where nodes on a layer are replicated on other layers with connections on layers being different and connections between layers being only between replica nodes. A queuing process acts on one network and uses resources from (this and) all other networks. An example of such a structure is given in \cite{aveklouris17mama,aveklourisstochastic}, where one layer is the electrical grid, another layer is the road network, and the queuing process is electrical vehicles queuing for parking and charging.

\noindent\textbf{Question 1.} What is the scaling limit of the queuing process in high-dimensions? How do simplifications affect this limit? Is it e.g.\ true that in dense settings, the scaling limit of the minimal spanning tree is the same as that on the complete network?

Quantifying the resilience of interacting networks is one of the main challenges in the field. As seen in nature \cite{bullmore09} or shown in research \cite{bianconi-MN}, adding layers improves resilience. From a queuing perspective however, adding layers adds dependencies, which may impact performance. In the other direction, as dependence may be an emergent property, controlling the queue can affect the structure of the network.

\noindent\textbf{Question 2.} What are the secondary effects of adaptive network structure on the performance of the queuing system? Vice versa, what effect does the control of the queuing process have on the resilience and dependence structure of the interacting network?

\section{Discussion}

Precise problem statements for questions such as those in Question 1 are easy and natural to formulate in queuing theory. As they are typically made concrete only in the context of a specific application, I have avoided details here. The problem statements are similar to those of queuing networks, with added complications due to the interdependencies. For example, in our work in \cite{aveklouris17mama,aveklourisstochastic} and follow-up papers, the capacity set can be non-polyhedral or even non-convex. Questions of stability, time-varying processes, multi-class arrival processes, and control can also be easily formulated, though technically their analysis is typically slightly more demanding than the classical cases and the outcomes oftentimes surprising.

Precise problem statements for questions such as those in Question 2 can be trickier. In one direction, the question may be stated as a question of the first type: increasing the resilience of the network may be formulated by some other fixed network structure, which can be taken as an input for the analysis of the queue. However, in the other direction such questions may require the description of the dependence structure (as emerging through the action of the queuing process on each layer) in terms of the network structure. Such translations do not  exist and it remains to be seen if the foundation that has been laid for the analysis of the structure of interacting networks is sufficient to capture the dependencies emerging from a queuing process acting on the network.

Ultimately, analysing queues in isolation (i.e.\ on one of the layers of the network) and thus ignoring the dependence structure of interacting networks can lead to significant performance estimate errors. For a simple two-layered network based on the machine-repair model, our results in \cite{DorsmanVlasiouBoxma} show that ignoring dependencies may lead to a 25\% estimation error.

\end{document}